# Using Terrestrial Laser Scanning, Unmanned Aerial Vehicles and Mixed Reality Methodologies for Digital Survey, 3D Modelling and Historical Recreation of Religious Heritage Monuments

**Aristeidis Zachos [1] and Christos-Nikolaos Anagnostopoulos[2,*]**

[1] Intelligent Systems lab, Cultural Technology and Communication department, University of the Aegean, Mytilene, Lesvos, Greece; cti20002@ct.aegean.gr

[2] Intelligent Systems lab, Cultural Technology and Communication department, University of the Aegean, Mytilene, Lesvos, Greece; canag@aegean.gr

* Correspondence: canag@aegean.gr; Tel.: (+30 22510 36624)

**Abstract:**

Preserving and safeguarding the Cultural Heritage (CH) of our world from unforeseen hazards should be viewed as a collective responsibility for humanity. Consequently, there is a growing imperative for targeted measures aimed at conserving, rejuvenating, and safeguarding historical assets that carry cultural significance. In recent times, Terrestrial Laser Scanning (TLS), Unmanned Aerial Vehicle (UAV) Photogrammetry, and applications in Mixed Reality (MR) have assumed a pivotal role in the mapping, recording, preservation, and promotion of Cultural Heritage. This article endeavors to present a comprehensive approach spanning from 3D surveying to the 3D representation and promotion of Religious Cultural Heritage, offering an overview of the applied methodologies.

Through the integration of TLS and UAV photogrammetry techniques, a comprehensive digital record of Panagia Ekatontapyliani, the adjoining Church of Agios Nikolaos, and the Baptistery, along with their wall paintings (hagiographies) and natural surroundings, has been obtained. This record serves as the foundation for historical documentation and recreation using the HBIM concept, paving the way for the development of diverse Mixed Reality applications. These applications aim to enhance the visibility, accessibility, and visitability of the Monument.

## 1. Introduction

The fundamental problems in the maintenance and preservation of historical buildings and monuments are detected in the lack of knowledge and financial resources, urban development and extension, which tend to take over and corrupt through various means or even eliminate all monuments and their area, and finally in the lack of a central planning by many states.

Despite the fact that a lot of monuments are protected by UNESCO, their number is still very small taking into consideration the cultural diversity that exists all around the world. Unfortunately, huge amounts of money are required, in order to cover the needs for all these monuments and objects, many years are needed in order funds to be found, as heavily maintenance, restoration and renovation actions must be performed. Consequently, until then, monuments deteriorate or even disappear.

Nowadays, in the digital era, the latest technologies of three-dimensional surveying and representation of objects can contribute to the preservation of cultural assets. In this context, Mixed Reality (MR) technology offers essential support for the digital restoration and promotion of monuments as means of historical information, as well as for the re-establishment of its initial historical value. This article seeks to provide a comprehensive perspective on the entire process, starting from 3D surveying and extending to the 3D representation and promotion of Religious Cultural Heritage, as well as to offer an overview of the implemented methodologies.

*1.1 Mixed reality*

Mixed Reality (MR) encompasses environments where real-world and virtual-world elements seamlessly coexist within the same visual field. The concept, initially defined in [1], spans the entirety of the virtuality continuum (VC), ranging from entirely real to entirely virtual environments, with augmented reality (AR) and augmented virtuality (AV) situated in between (see Figure 1). Implementation of this continuum is achievable through devices such as Head Mounted Displays (HMDs) or mobile devices, seamlessly integrating real-world scenes with digital information channels.

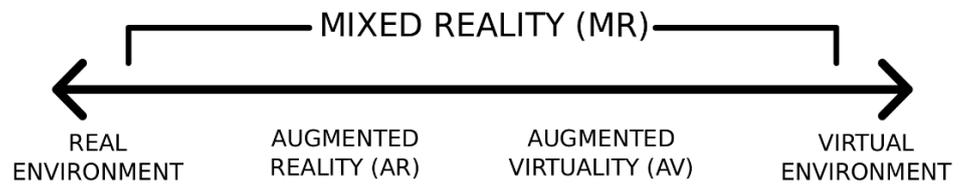

**Figure 1.** The Mixed Reality continuum as proposed by Milgram and Kishino[1].

Augmented Reality and Augmented Virtuality (AV) are the two predominant methods for developing MR applications. AV integrates real-world elements into the Virtual Environment (VE), while AR functions in the opposite direction, introducing digital objects into the Real Environment (RE). Virtual Reality (VR) marks the extreme end of the continuum, showcasing fully synthetic worlds that fully immerse users. In AR applications, the Real Environment is typically presented through a mobile device camera or HMD, with digital objects overlaid on the RE using computer vision algorithms for fiducial marker detection or relative sensors (GPS, accelerometers, etc.) for accurate positioning in the scene.

Advanced MR applications often eschew markers (marker-less AR), presenting a challenge that has garnered considerable attention. Some high-end HMD platforms incorporate eye tracking to enhance user input in AV or VR applications. Beyond visual inputs, popular VR and AR development platforms leverage sophisticated sensors for room-scale applications and may include joysticks for comprehensive hand or full-body support.

Photogrammetry and terrestrial laser scanning (TLS) are pivotal techniques in supplying data to MR applications. These methods produce accurate models and digital outputs, contributing realistic inputs to MR environments. While TLS produces highly accurate models through an automated process, its drawback lies in its high cost, requiring specialized and expensive equipment and expertise.

*1.2 Photogrammetry and UAV photogrammetry*

The term "photogrammetry" derives from "photography," a compound of "photo" (from the Greek term meaning light) and "graphy" (from the Greek term meaning recording or writing). To further define "photogrammetry," the Greek term "metron" (denoting a device for measurement) is added. Essentially, photogrammetry can be understood as a technology that measures objects through the recording of light. The American Society for Photogrammetry and Remote Sensing offers a broader definition, describing it as "the art, science, and technology of obtaining reliable information about physical objects and the environment through processes of recording, measuring, and interpreting images and patterns of electromagnetic radiant energy and other phenomena" [2]. In contemporary times, it stands out as one of the most widely used techniques for acquiring data and generating 3D models.

Photogrammetry proves to be a cost-effective approach for acquiring large-scale digital data. When surveying cultural or religious monuments, ground-level images are the most convenient and readily accessible data. However, the limited view in each photo, coupled with occlusions, poses challenges in surveying expansive areas. Consequently, aerial surveys (or Unmanned Autonomous Vehicle Photogrammetry - UAV Photogrammetry) are frequently conducted using drones (UAVs) equipped with high spatial resolution and high-quality cameras that capture nadir and/or oblique images. Oblique images, taken at a tilt angle, enhance visibility of monument facades, buildings, or other vertical structures (e.g., fortification walls). Skilled UAV pilots carry out low-altitude planned or manually operated flights to acquire oblique images. Processing an appropriate image set using suitable software and algorithms can yield highly accurate 3D models. To ensure accuracy, it is crucial to capture images within the context of a well-planned surveying UAV flight.

In terms of software, Structure from Motion (SfM) stands out as the most popular methodology for both terrestrial and aerial photogrammetry [3]. Numerous software options based on SfM are available, including open-source choices (VisualSFM, AliceVision, OpenMVS, COLMAP, and Regard3D) and commercial ones (Agisoft Metashape/PhotoScan, 3DF Zephyr, Reality Capture, Autodesk Recap Photo, and Context Capture).

*1.3 Terrestrial Laser Scanning*

Terrestrial Laser Scanners (TLSs) have the capability to generate high-quality 3D models for large-scale natural formations, archaeological sites, monuments, structures, or objects. A significant advantage lies in their ability to record accurate and non-destructive data within a relatively short surveying time. However, due to limitations such as space constraints, occlusion, and the need for a comprehensive survey, multiple scanning positions are required, carefully selected to address these challenges. The culmination of multiple scans results in a final 3D point cloud model of the surveyed scene, a process that will be elaborated on in the subsequent section of this paper. This approach helps overcome issues like occlusions, scan range limitations, and restricted scan views.

The configuration of scanning positions, orientations, and parameters is a critical aspect, aiming to reconstruct the point cloud and 3D model with optimal quality and a minimal number of scans. The utilization of TLS for heritage documentation has witnessed a surge in popularity.

The underlying technology of TLS is LiDAR (Light Detection and Ranging), a technique employed to determine the distance of each object point from the lens. LiDAR operates by emitting a beam or pulse series of highly collimated, directional, coherent, and in-phase electromagnetic radiation. Upon receiving the reflected light from an object's surface, the system calculates the range based on flight time and captures the reflectivity of the surface.

Two distinct methods of range determination exist: phase and pulse [4]. The former is more accurate but has a limited range, while the latter can measure over a greater distance. Consequently, the pulse method is commonly implemented in TLS used for civil construction measurements.

The combination of TLS and UAV photogrammetry techniques yields comprehensive 3D products for complex objects. The continual advancement of computational power and the evolution of sophisticated hardware for data acquisition and processing offer substantial potential for accurate modeling. These two technologies complement each other by addressing potential occlusions from various perspectives, such as roofs or the tops of buildings and monuments.

## 2. Review of related literature
*2.1 Review on AR/VR for cultural and religious heritage*

As far as the digital reconstruction of religious cultural heritage is concerned, a range of scholarly works has been published covering: a) the representation of monumental/mural paintings and frescos, b) reconstructions and visualizations of temple/church architecture, c) Historical Building Information Models (HBIM), and d) the development of accurate 3D Digital Twins platforms for monuments.

In the context of monumental paintings and frescos, Petrova et al. outlined a virtual reconstruction workflow for the restoration of paintings in the Church of the Transfiguration of Our Saviour on Nereditsa Hill [5]. Girbacia et al. utilized Augmented Reality technologies to enable the virtual restoration of heritage objects, employing geolocation and computer vision [6]. Soto-Martin et al. recently presented a digital reconstruction of a historical church as an interactive and immersive experience in virtual reality, with a focus on using digital imaging tools for the recovery of deteriorated mural paintings [7].

Most case studies in this domain have centered on the digitalization of religious cultural heritage and its exploration through Virtual Reality (VR) in an immersive manner. An initial exploration of temple/church architectural reconstruction using VR applications was presented in [8]. Recent works have delved into workflows connecting photogrammetry and VR for digital religious heritage, as showcased in [9-12]. These papers discuss the interactive and immersive experience of VR in the digital reconstruction of historical religious buildings and monuments, often utilizing multi-source 3D data, a combination of photogrammetric and terrestrial laser scanning data.

Heritage/Historic Building Information Modeling (HBIM) is emerging as a novel approach for creating intelligent 3D models of cultural heritage monuments. This method incorporates historical archive information and digital data, such as restoration studies, architectural and conservation practices, and cultural heritage documentation. The potential of HBIM to contribute to understanding complex relationships between tangible and intangible religious heritage is highlighted in [16-18]. Bacci et al. [19] demonstrate a study incorporating 3D architectural survey, critical analysis of historical sources, and a final conservation proposal, while [20] presents a holistic procedure for building the BIM model directly from point clouds for the architectural heritage evolution of Saint Jerome's Church in Baza.

In recent years, the application of Digital Twin (DT) principles to support site managers in the preventive conservation of assets has gained prominence. Digital Twins, HBIM, and Mixed Reality are integrated in the case study of St. Francis of Assisi Church in the Pampulha Modern Ensemble in Belo Horizonte, Brazil [23]. The comprehensive pipeline established by Dezen-Kempter et al. covers all steps from 3D point clouds to HBIM and then to a Digital Twin platform, providing Augmented Reality products.

## 2.2 Review on photogrammetry and 3d scanning for religious cultural heritage

Numerous research initiatives have explored diverse methodologies for generating 3D models through photogrammetry and terrestrial 3D scanning. Notable examples include studies conducted in the Church of San Francisco, the Church of A Coruña [24], the Cathedral of Notre-Dame des Amiens [25], and the Cathedral of Jaén [26]. Additionally, in [27-29], the authors proposed a cost-effective approach for 3D registration and point cloud extraction from ordinary 2D images, employing the structure-from-motion (SfM) method and dense image stereo-matching for archaeological heritage sites. The results, whether in 2D or 3D, demonstrate high accuracy and integration potential with archaeological excavation plans. Numerous studies illustrate the application of 3D scanning, photogrammetry, and close-range cameras for monitoring cultural heritage sites [30,31]. In [32], a project is outlined for the reproduction of sculptures, namely Saint Teresa de Jesús and Christ Tied to the Column, situated in the Convent of Santa Teresa in Ávila, Spain. Several studies have also integrated the use of close-range Unmanned Aerial Vehicles (UAVs) for 3D documentation [33-37].

## 3. Methodology for terrestrial scanning and aerial photogrammetry. Case study: Panagia Ekatontapylianni

### 3.1 History of the monument

The case study of this article is the complex of the Holy Church of Panagia Ekatontapiliani, Saint Nikolaos and Baptistery, which are all located in the island of Paros. The complex of the Church was founded in the ruins of a large ancient gymnasium in Paros (Figure 2, left). It has re-mained almost intact since the era of its final construction (6th century) and, at the same time, it is still seen as one of the greatest Early- Christian monuments along with the Church of Saint Demetrios and the Church of Virgin Mary Acheiropoietos in Thessaloniki.

The story of this magnificent monument dates back to the 4th century and more specifically, right after 313 (AD) and the decree on tolerance for the religion (Edict of Milan) from the present Holy Church of Saint Nikolaos. This Holy Church was dedicated to Virgin Mary before it was renamed Saint Nikolaos, from the erection of the central Holy Church of Virgin Mary and the Baptistery. First, this Holy Church used to have a wooden roof, but after being destroyed by a fire, it was then rebuilt based on the old floor plans with a dome [38].

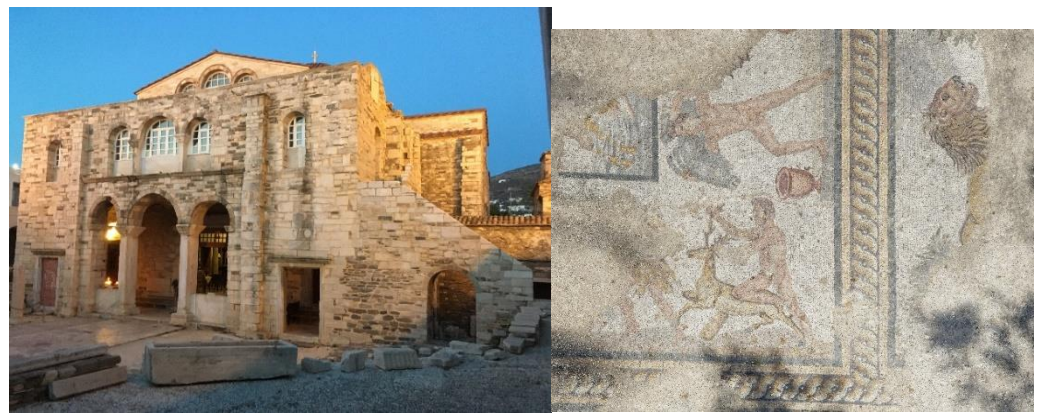

**Figure 2.** (Left) front view of the Monument, (Right) part of the mosaic depicting the labors of Hercules.

During the restoration processes in 1960 and the exploratory excavations, a mosaic with a depiction of the labours of Hercules was discovered, which belonged to an ancient gymnasium that existed before the Holy Church of Virgin Mary (Figure 2, right). This mosaic was found under the floor of the Holy Church and was transferred several tens of meters to the southeast, in the forecourt of the archaeological museum of Paros. This mosaic will be a Mixed Reality case study in this paper.

*3.2 Preparatory stage of 3D Terrestrial Laser Scanning (TLS)*

Prior the actual TLS survey, a specific planning was conducted to ensure that the coverage of the monument from outside is complete. The preliminary phase was carried out ex-situ, employing a methodology devised in our laboratory (i-lab.aegean.gr). This method serves as a reconnaissance exercise, guaranteeing: a) the optimization of Terrestrial Laser Scanner positions and the scanner's field of view to uphold point cloud quality while minimizing scanning positions, and b) adherence to quality constraints regarding the point cloud. These constraints encompass allowable incidence angles, minimum/maximum scan range, field of view, potential noise factors (such as the presence of vegetation, vehicles, people, visitors, fencing, or shadows).This methodology is based on coarse top-views acquired from google maps or specific plans that are possible available and is briefly depicted in Figure 3, using Unity Game Engine (https://unity.com/).

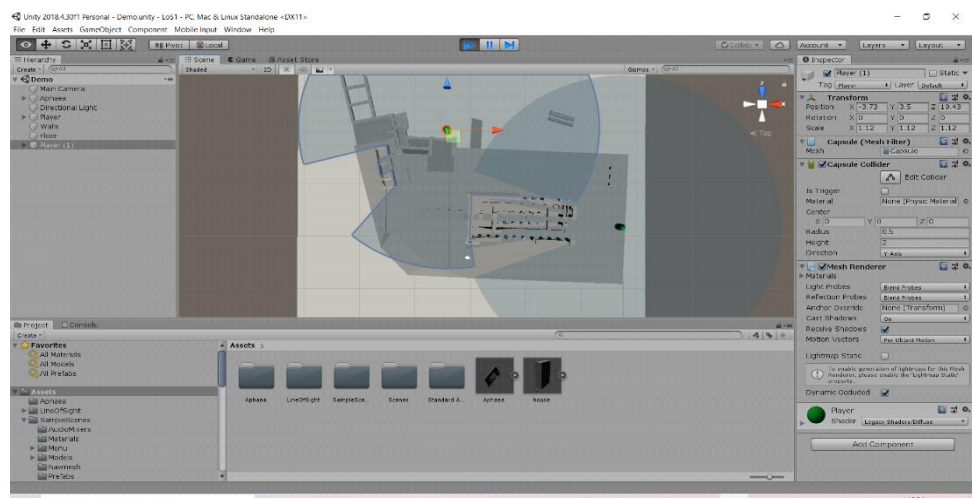

**Figure 3.** Virtual positioning for initial optimization of TLS scanner in the top-view of the monument (screen shot from Unity).

3.2.2 Survey – specifications – material and methods

Considering the technical specifications of the Terrestrial Laser Scanner (TLS) and the specific characteristics of Panagia Ekatontapyliani, a carefully planned set of 163 scanner positions was acquired. This configuration aimed to achieve maximum coverage of the monument while ensuring the production of a point cloud with a minimum spacing of less than 1.5 cm for the entire survey. Each scanning position was meticulously executed to provide substantial overlap (>20%) between consecutive laser scans, facilitated by the use of auxiliary targets.

The TLS survey was carried out using a FARO Focus3D phase-shift laser scanner, and a total of 163 laser scans were performed. The scanning settings included: i. a scan area of 360° horizontally × 305° vertically, ii. ¼ resolution, acquiring point clouds at a spacing of 6.13 mm/10 m, thereby achieving less than 1.5 cm for the entire monument scenario, and iii. digital compass and inclinometer measurements activated.

Table 1 outlines the specifications of the scanner. In addition to that, in order to cover missing areas due to occlusions or lack of accessibility (e.g. roof of the monument), the method of photogrammetry by air was used, as presented in the following section.

**Table 1.** Technical specifications of the 3D Terrestrial Laser Scanner (TLS).

| TLS Specification Type | Value |
|---|---|
| Max measurement speed | 976.000 pts/sec |
| Ranging error | ±2 mm |
| Color resolution | Up to 70 megapixel color |
| Field of view | 305°vertical/360° horizontal |
| Step size | 0.009 (40,960 3D-pixel on 360°) vertical/0.009 (40,960 3D-pixel on 360°) horizontal |
| Laser Class | Laser Class 1 |
| Wave length | 1550nm |
| Parallax | Minimized due to co-axial design |

The survey process started with the on-site inspection and analysis of the monument. A circular route was decided and then applied outside the monument with inter-related stops, aiming at depicting the external outline as precisely as possible. Next, two routes were designed to the interior of the monument; one on the top of the gynaikonitis (i.e women area) and a second one leading to the baptisterion (baptistery), the central part of the temple and the church of Saint Nikolaos (Figure 4).

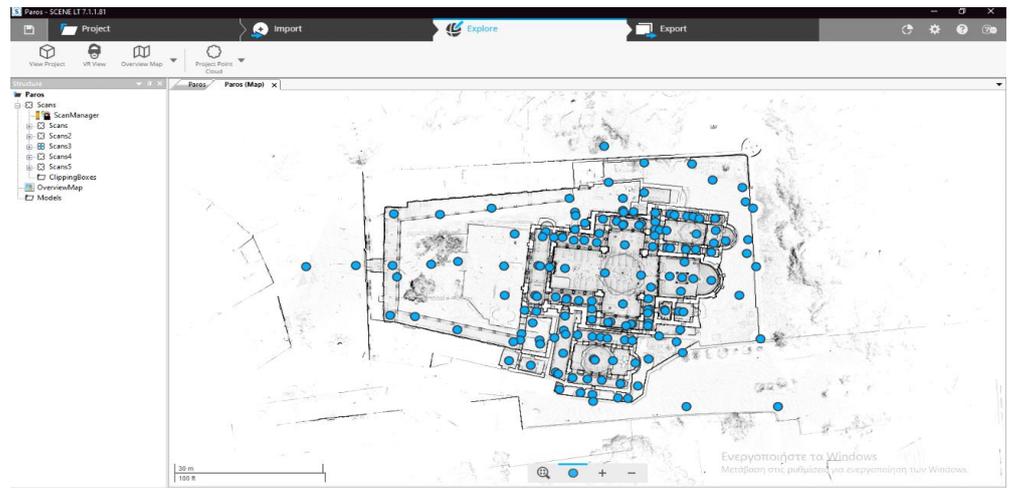

**Figure 4.** Part of 3d scanner scanning stations (depicted as blue dots).

Scanner positions were also selected, with a view to covering and measuring the route in the best possible way, as well as avoiding various obstacles that might intervene between the scanner and the object. Moreover, due to the size of the monument, the number of scans should also be constrained, in order the final outcome to be managed more easily by the editing software. In most scans, the maximum distance for scanning objects did not exceed 10-15 meters, while the overlapping between the scans reached at least 20%.

For the optimization of the registration process of consecutive scans, at least 3 spherical targets of 15 cm diameter were placed between each pair of scans. The distance between the scanner-object did not exceed 10-15m at all scans. This led to the best possible deliverables, since with this configuration the laser beam of the scanner has the highest precision possible to be achieved.

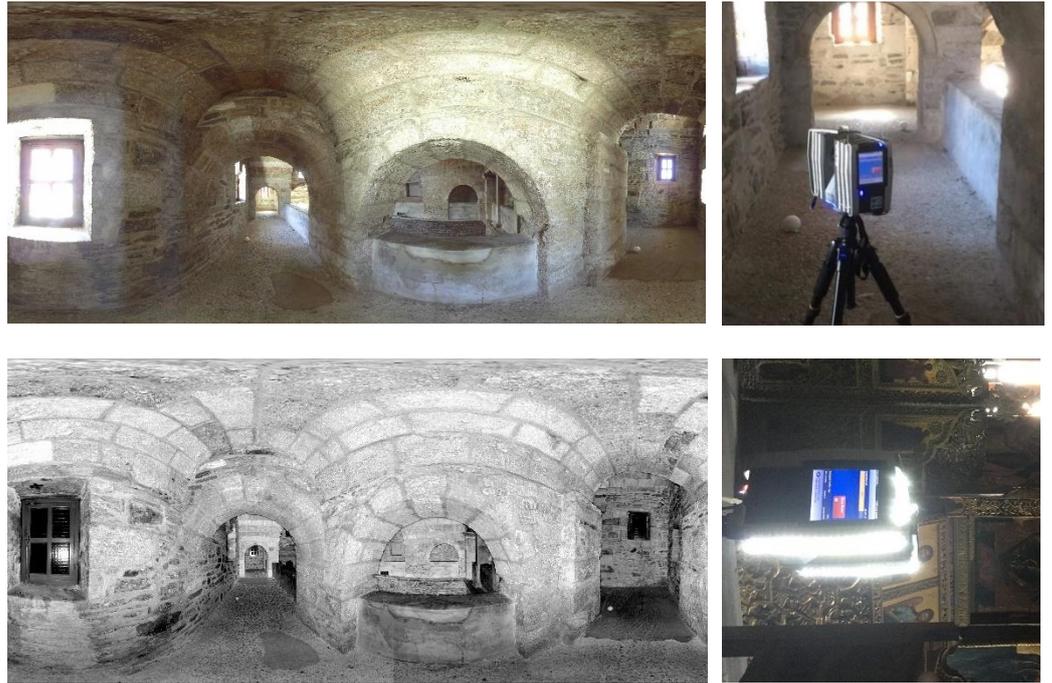

**Figure 5.** (Up) scan with RGB color, (bottom) scan with infrared. Auxiliary spheres are noticeable.

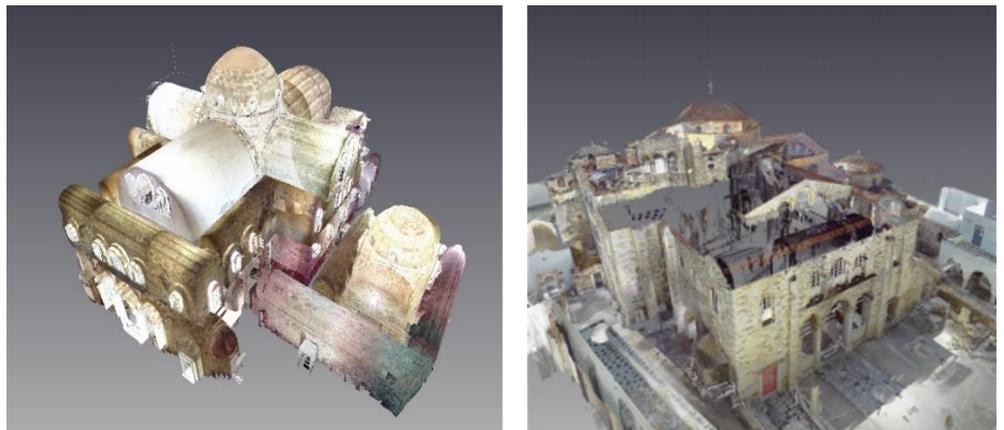

**Figure 2.** Left: The overall TLS model (Left: interior, Right: exterior).

During point cloud acquisition in every scanning position, infrared and color photos were also collected, using the built-in camera of the scanner, to capture color and surface texture. Moreover, an external led lighting placed on the top of the laser scanner was used, to properly capture surface details in the interior of the church, where illumination conditions were uncontrollable, such as the sanctuaries and the gynaikonitis (see Figure 5). At the same time, visual checking of all scans was also performed, to ensure that all individual point clouds had received the appropriate data. The overall TLS 3D point cloud on the interior and the exterior of the Monument complex is shown in Figure 6. The point cloud registration delivers a unified total point cloud in a single global coordination/reference system for both the interior and the exterior of the Monument. During the final check, the maximum error reached 7.3mm.

*3.3 Photogrammetry with Unmanned Aerial Vehicle (UAV)*

3.3.1 Procedure and methodology

We used the Software from Motion (SfM) photogrammetric processing, which converts an appropriate set of images into various digital products, vector and bitmap images, digital orthophotos, 3D models and Digital Elevation Maps (DEM). The overall processing can be summarized in the following steps:

- automatic photo orientation by locating homologous and scale invariant features in all images by identifying their descriptors
- Solving aerotriangulation, performing self-calibration and eliminating measurement errors (using Ground Control Points -GCP). Aerotriangulation is a method employed to ascertain the coordinates of terrain points using aerial photographs. The primary objective is to increase the density of a geodetic network, providing aerial photographs with the essential control points required for the compilation of topographic maps.
- Automatic production of cloud points in the whole area through local/global matching algorithms
- Noise elimination and creation of digital terrain model (surface or ground)
- Creation of a three-dimensional mesh object model (3D triangulation)
- Calculation and refinement of a texture image to be projected in the model

To achieve this objective, an Unmanned Aerial System (UAS), specifically the Phantom 4 Pro, was employed to capture both nadir and oblique images using a fc6310 20MP - 1-inch CMOS camera with a fixed lens of 24mm focal length. The aerial survey of the monument involved flights at various heights to capture the architectural details optimally (an indicative flight plan is shown in Figure 6). Survey missions were conducted at flight heights ranging from 20 to 30 meters, resulting in a mean ground resolution of 5.93mm/pixel. Flight planning ensured an 80% overlap in-track (flight direction) and an 80% cross-track (side lap). The flights, with a speed of 1.5m/sec and a maximum duration of 18 minutes, were executed autonomously and semi-autonomously, with continuous operator monitoring to ensure adherence to the planned procedures.

Given the complexity and structural state of the monument, a combination of two distinct flight plans was deemed necessary: one with a nadiral camera configuration and the other with an oblique configuration. For nadir images, the UAV pilot defined the survey area around the monument on the application basemap. In contrast, for oblique images, a point of interest was set at the center of the monument, and the drone followed a radius of approximately 15m.

A total of 708 aerial images were acquired, and after a quality check, 707 of them were utilized in the Structure from Motion (SfM) process. Among these images, 409 were nadir and 298 were oblique. The aerial survey predominantly occurred around sunrise and sunset to mitigate strong reflections caused by sunlight. Additionally, a small set of images was manually acquired during flights to target challenging locations of the monument, ensuring comprehensive coverage and addressing potential blind spots.

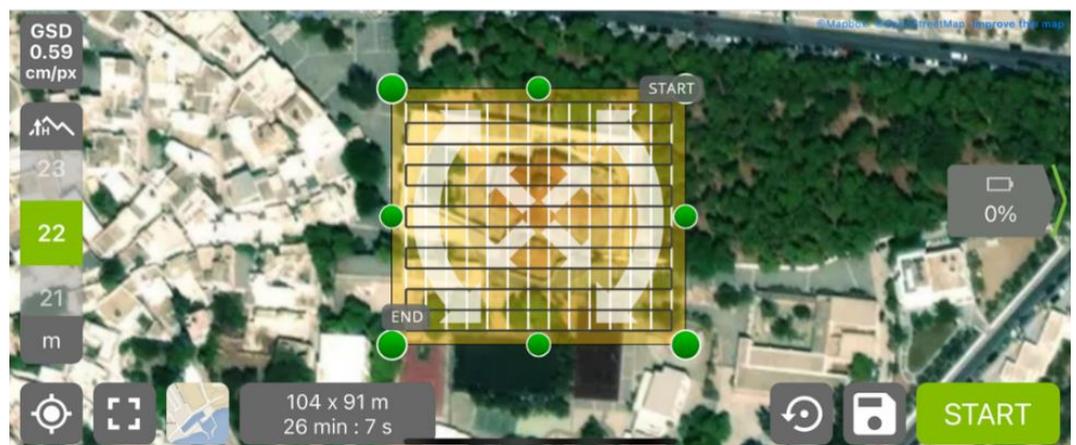

**Figure 3.** Screenshot of the initial stage of one autonomous flight plan.

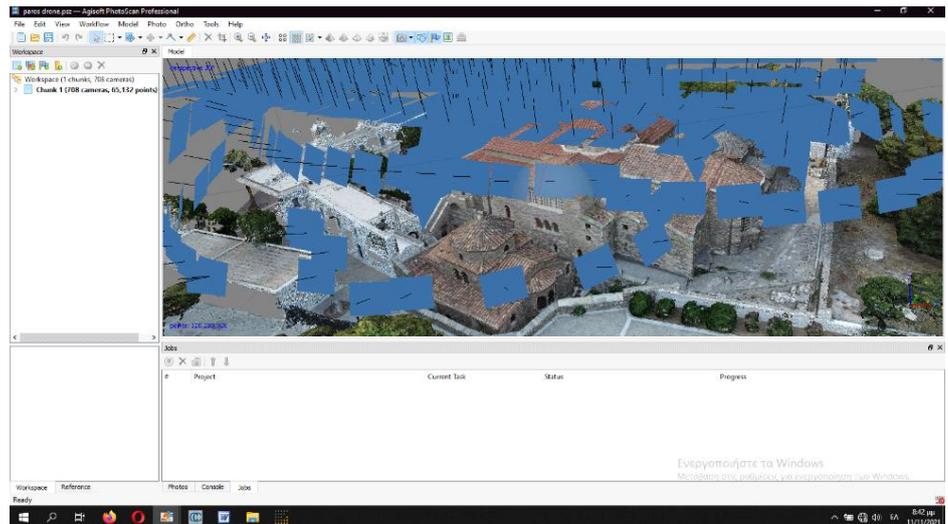

**Figure 4.** Image alignment in Photoscan/Metashape.

3.3.2 UAV images – photogrammetry and post-processing

The image processing was performed by Agisoft Photoscan/Metashape, and the parameters of the workflow are shown is Table 2. Global Positioning System (GPS) information, as well as the successive photo annotation contributed to a faster and more precise processing of the resulting 3D point cloud model. Ground Control Points (GCPs) were also used, which are points on the ground with known coordinates in the referenced spatial coordinate system. Their coordinates are obtained with GNSS rover in RTK mode. Specifically, for the current survey five (5) GCPs were used for the optimum orientation and placement of aerial photographs in the spatial coordinate system, as well as for the successful production of georeferenced metric, 3D point cloud (see Figure 8) and orthophotos.

**Table 2.** UAV photogrammetry parameters.

| Parameter | Value | Parameter | Value |
| --- | --- | --- | --- |
| Number of images | 708 | Camera stations | 707 |
| Flying altitude | 24.9 | Tie Points | 65,132 |
| Ground resolution | 5.93mm/pixel | Projections | 399,097 |
| Coverage area | 8.46e+03m2 | Reprojection error | 1.21 pixels |
| Image Resolution | 5472x3078 pixels | Pixel size | 2.53x2.53 μm |

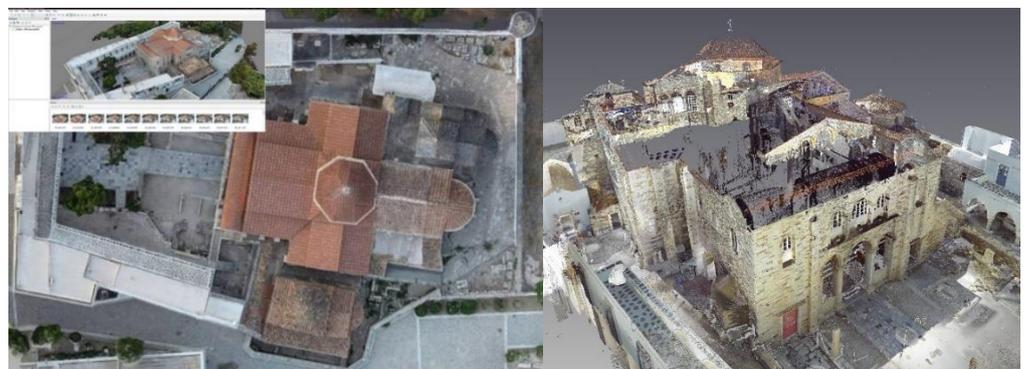

**Figure 5.** (Left) final 3d point cloud from UAV photogrammetry, (Right) screenshot of the final TLS 3D point cloud.

*3.4 Metaprocessing – fusion of TLS and UAV point clouds*

After the collection of all point clouds from both techniques (TLS and UAV photogrammetry), a demanding and meticulous office work follows, where the registration of point clouds acquired from photogrammetry and terrestrial scanning is carried out semi-automatically or automatically using specialized software (i.e. Faro Scene). The workflow is depicted in Figure 8, and it can be summarized as follows:

1. Registration of point clouds from UAV Photogrammetry and TLS to one complete point cloud and mesh (see Figure 9, left). The registration was performed with the use of CloudCompare software, taking into consideration all assistive information from the reflective spheres and the CGPs.
2. Delivery of 3D elevation maps (Figure 9, right) and orthophotos (Figure 10)
3. Files preparation to final editable forms
4. Delivery of Mixed Reality and Digital Products as presented in Section 4

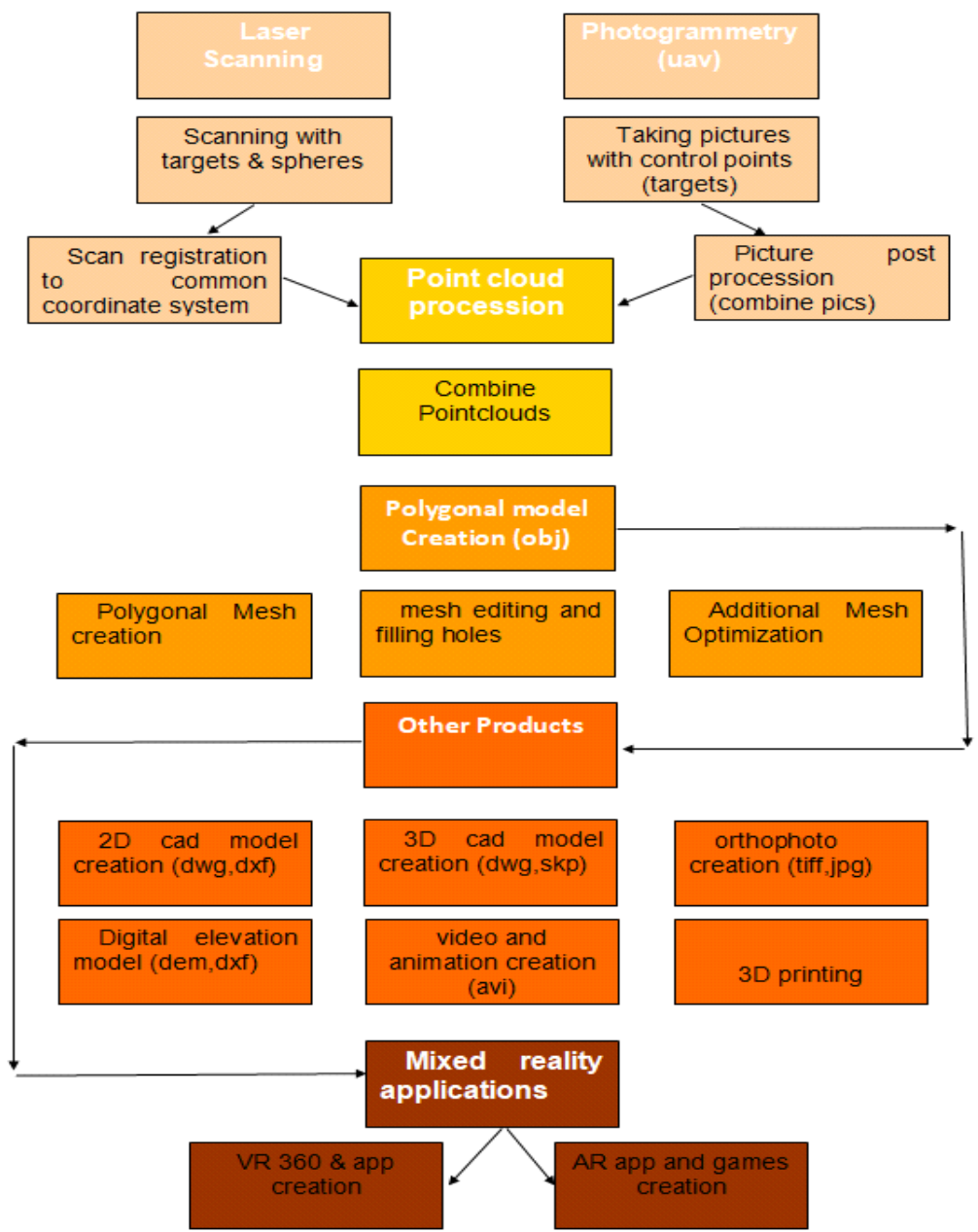

**Figure 6.** Metaprocessing (fusion of scanning results) and digital products delivery.

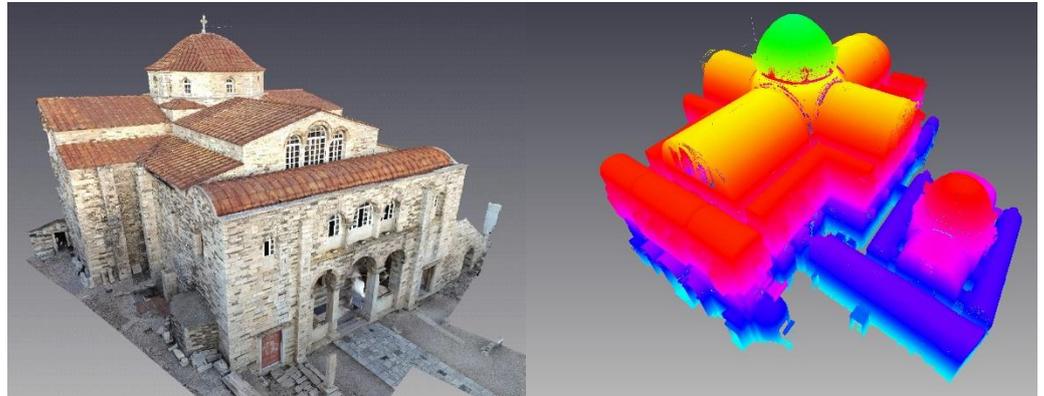

**Figure 9.** (Left) final 3D mesh after the final registration, (Right) 3D elevation map.

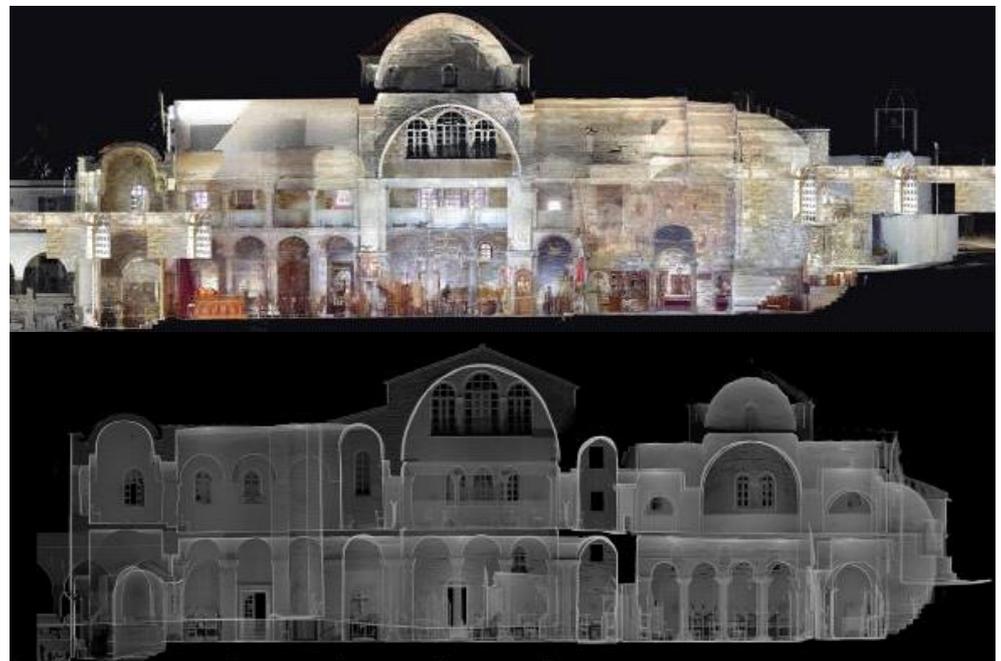

**Figure 7.** Orthophotos of the central section with and without color information.

## 4. Digital products and Mixed Reality applications

### 4.1 Representation in HBIM

Building Information Modeling (BIM) has enabled the modeling of historic structures by depicting the entire life cycle of buildings, starting from architectural plans and drawings to the completed construction state, including the materials used. BIM establishes relationships among products, spaces, systems, and sequences, preventing potential errors in development or construction stages. Derived from this, Heritage (or Historical) Building Information Modeling (HBIM) was introduced to cater to the specific needs of heritage conservation and safeguarding.

In this study, the Sketchup software was employed to create an initial HBIM model for Panagia Ekatontapyliani. Given that HBIM was not the primary focus of our survey, the model was developed with a focus on generating the 3D architectural model, offering limited input of historical or constructional metadata. For illustrative purposes, a screenshot of the HBIM model is presented in Figure 11. The HBIM model attained the 2nd Level of Detail (LoD2), encompassing the main architectural details, primary structural elements, and the penetrations of the monument (such as floors, columns, beams, and basic structures of doors, openings, and windows). It should be noted here that the HBIM includes a wooden roof instead of a dome (as the Church has now). This is due to the fact that during the first establishment of the Holy Church in the 4th century, it was constructed with a wooden roof similar to the one of another Early Christian Church, the Church of Saint Demetrious in Thessaloniki. The present architectural form with the dome, was developed in an earlier period (6th century), after the destruction of the monument due to fire. Therefore, the wooden dome is displayed as a historical representation in the HBIM model as depicted.

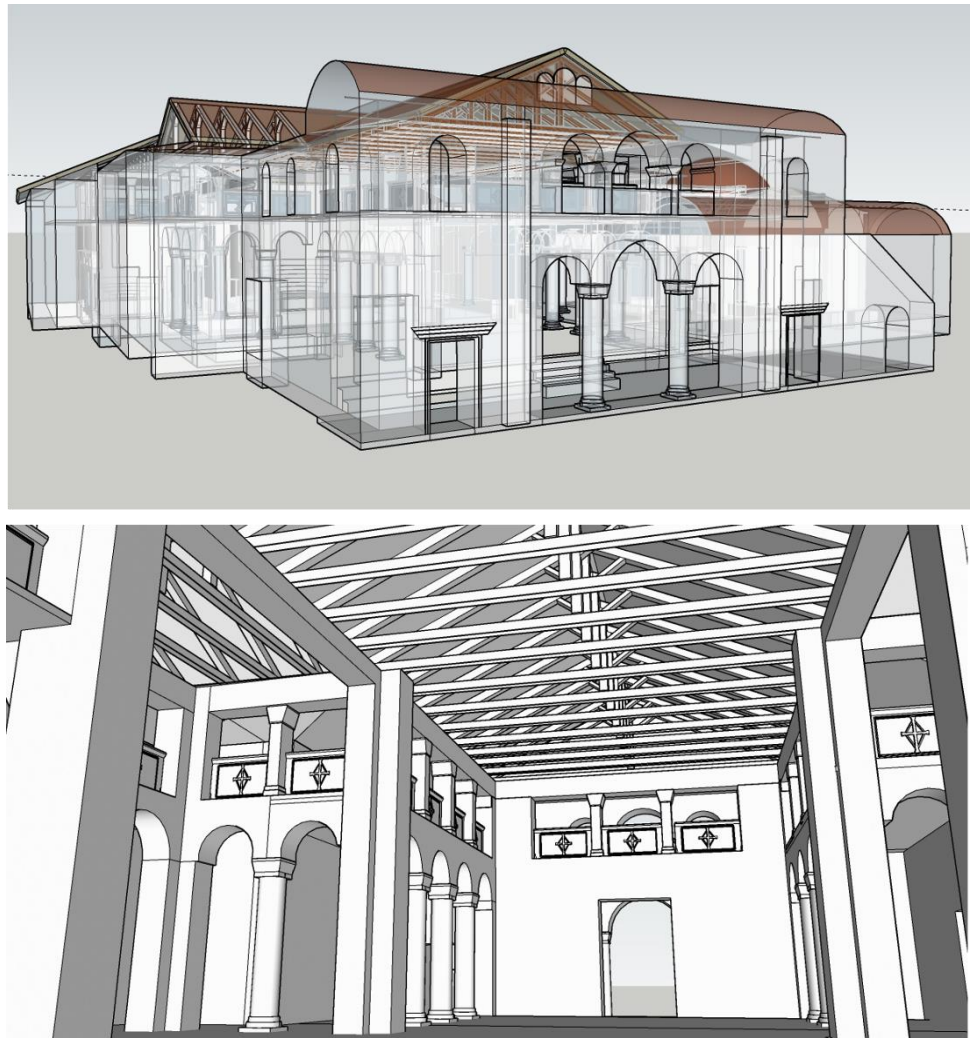

**Figure 11.** Screenshots from the HBIM model is Sketchup (LoD2). Up: Exterior view, Down: Interior view. Historical representation of the first period of the Holy Church (before 6th century A.C.).

*4.2 Virtual reality application*

Based on the monument survey, the Virtual Reality (VR) products include spherical 360 images to build paths of virtual navigation/tours using the appropriate software (PanoTour). Virtual tours are available to laptops/PCs, as well as mobile devices, and virtual visitors may visit rooms and places that are not normally accessible, such as the Ciborium of the Altar or the Synthronon. Some indicative unfolded 360 spheres are shown in various specifications in Figures 12 and 13 as well as in [39]. Figure 14 shows how the user may experience as an avatar a 3D time machine tour of the overall complex of the monument in its early period.

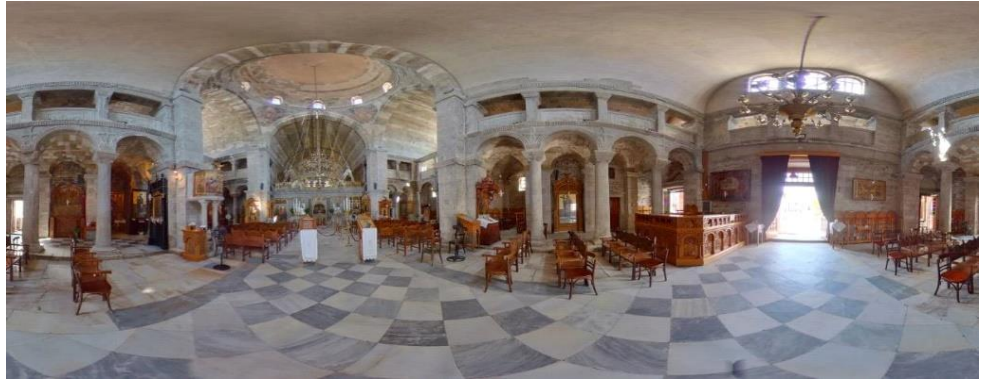

**Figure 12.** Unfolded 360 spherical image inside the Holy Church.

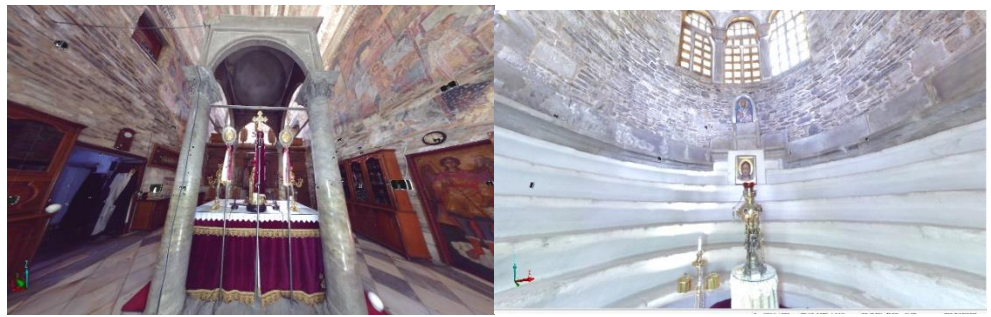

**Figure 13.** Unfolded 360 spherical image of Synthronon and Kivorion.

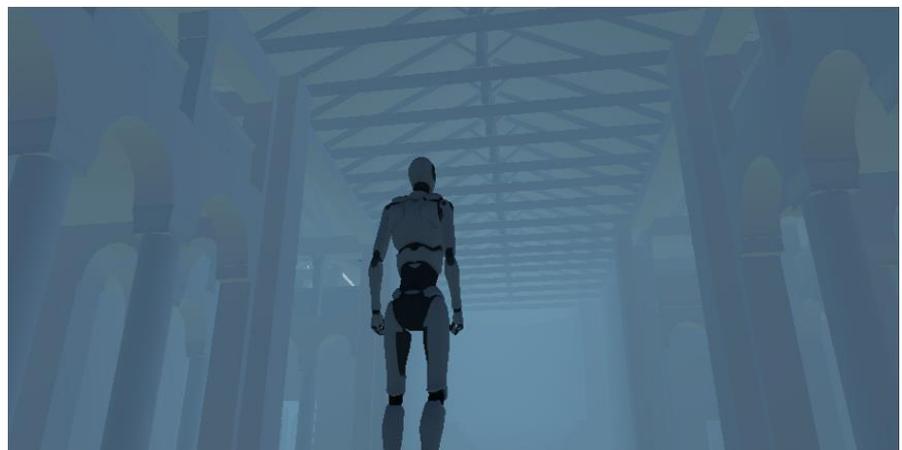

**Figure 14.** Configuration of 360 image split screen to be projected in a VR mask or VR mobile device.

### 4.3 *Augmented reality*

An Augmented Reality application proposed in this paper is implemented using a simple smart-phone device, through which users can see Hercules's mosaic in their initial place simply by turning their cell phone towards the floor of the Church, as shown in Figure 17.

This marker-based application was build using the Unity Game Engine (https://unity.com/) integrated with the Vuforia Augmented Reality platform. Unity is a game engine capable of developing computer, console or mobile game. The Vuforia SDK (https://developer.vuforia.com/) stands as an Augmented Reality (AR) software development kit designed for mobile devices and introduced by Qualcomm. Leveraging computer vision technology, it facilitates the real-time recognition and capture of planar images or 3D objects. Developers can utilize Vuforia to overlay virtual objects onto the camera viewfinder, adjusting the position of these objects against the camera's background. Vuforia SDK provides support for various types of 2D and 3D objects, encompassing multiple target configurations and images featuring fewer symbol and frame tags. The result is mobile app that identifies a marker (QR-code) placed in the floor of the monument and projects a 2D image of the ancient mosaic in the appropriate place in the floor through the mobile device.

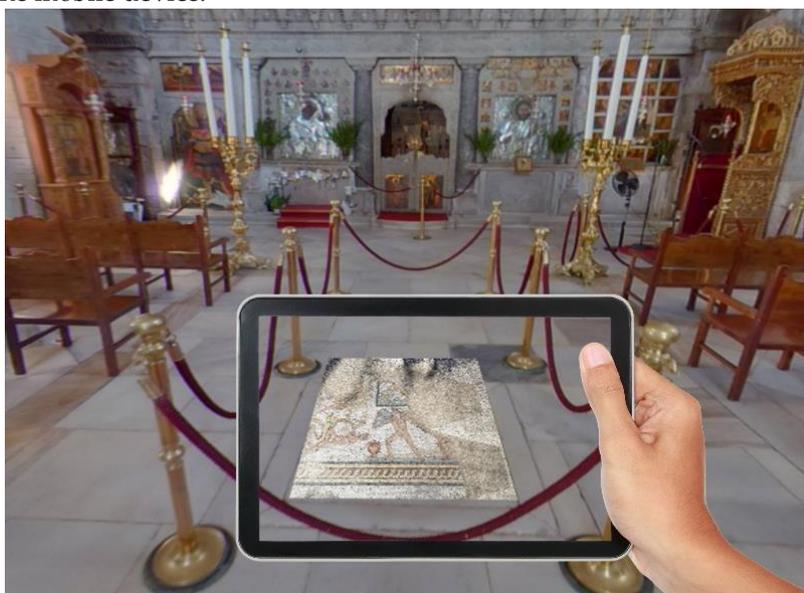

**Figure 15.** Augmented Reality of the mosaic of Hercules. A marker is placed in an suitable place in the floor.

A second AR application could project a 3d model of a wooden roof in the screen of the mobile device instead of the dome, turning it mobile phone upwards. This can be suggested, since during the first establishment of the Holy Church in the 4th century, it was constructed with a wooden roof similar to the one of another Early Christian Church, the Church of Saint Demetrious in Thessaloniki. The present architectural form with the dome, was developed in a latter period (6th century), after the destruction of the monument due to fire. Unfortunately, such an AR application is very hard to be implemented due to: a) the existence of various objects (e.g. huge chandeliers) that occlude the target scene (i.e. dome), b) the height of the dome from the ground and c) the overall architectural complexity of the monument. Therefore, the wooden dome is displayed as a historical representation in the HBIM model as depicted in Figure 18.

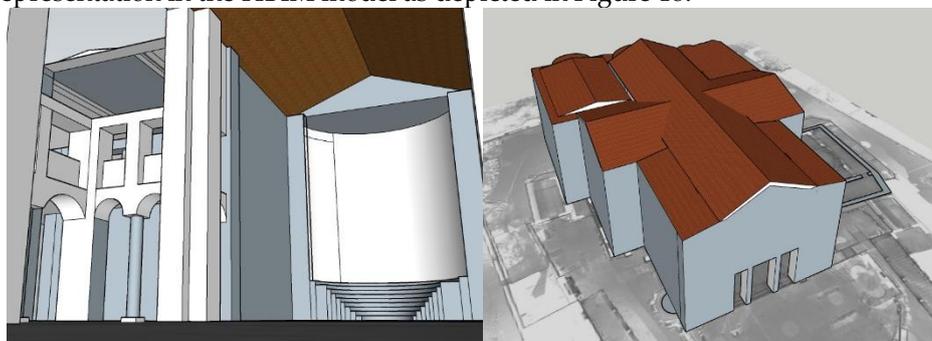

**Figure 8.** The wooden roof as displayed in Historical BIM platform (inside and outside view) in LoD2.

*4.4 3D printed model*

Additive 3D manufacturing (3D printing) introduces new eras regarding the exhibition and promotion of Cultural and Religious Heritage. Specifically, it enhances the understanding, the communication channels and the accessibility of historical monuments as it enables tactile access to information in cases that this not possible (e.g., visually impaired people). To this end, people with vision deficiencies may touch and feel printed replicas of monuments or relics of Religious Heritage in various Levels of Details (LODs). Figure 14 displays the 3D printing stages of the STL model of the Holy Church of Panagia Ekatontapyliani using a 3D-printing software, is depicted, along with the respective printed layers and the supporting materials.

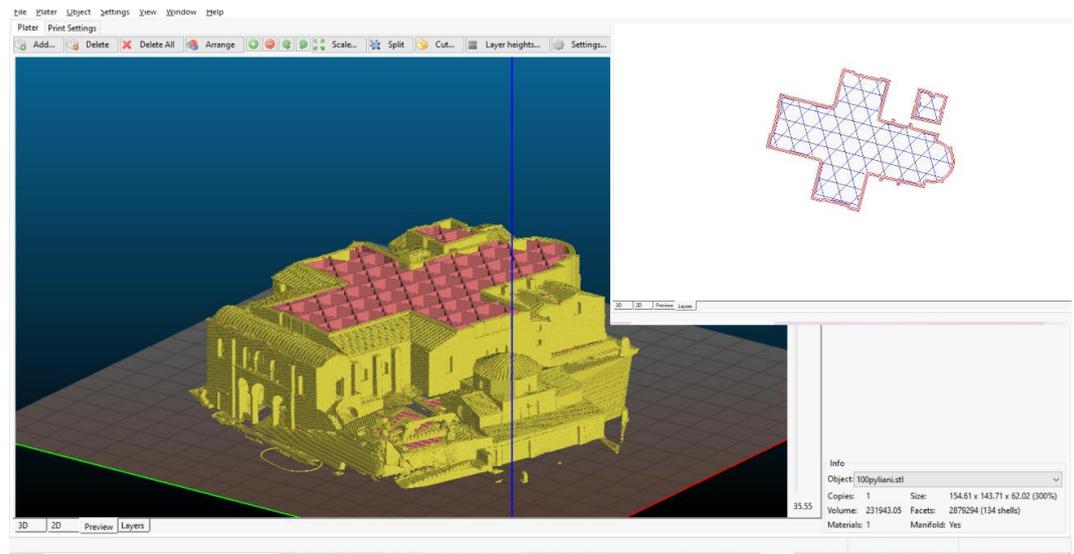

**Figure 9.** 3D printing process of STL model of the monument.

## 5. Conclusions

### 5.1 Summary

The aim of this paper is to present a holistic methodology for surveying Monuments of Religious and Cultural Heritage, highlighting the numerous capabilities and digital products that new technologies offer nowadays. Digitization of our cultural heritage is extremely useful for scientists and experts, such as archaeologists, architects, surveyors, conservators, museologists etc, since three-dimensional models, ortho-images, facades, floor, and section plans can be produced. In addition, the general public may be benefitted, since the delivered digital products can be exploited for touristic and/or educational purposes using Mixed Reality methodologies and 3D printing. The main goal of similar research and surveys is to concentrate on the preservation of Cultural and Religious Heritage at risk, the need to reverse engineer similar architectural models and deliver Digital Twin models, which may increase the immersive learning, visualization, and promotion of our Heritage.

Since preserving our world Cultural Heritage (CH) from the unpredictability of hazards must be seen as a duty towards whole humanity, the responsibility lies on the appropriate acts for preserving the historical memory of everything holds a cultural value. The added value of focusing on TLS and UAV surveying methods and workflows in the Cultural/Religious Heritage safeguarding, promotion and restoration, is that a direct connection and an architectural accuracy documentation with the real CH case study is provided thanks to the reality modelling approach, which is typical of the 3D metric survey operations. The mutual co-operation and synergy between 3D survey methods, cultural studies and computer science has strongly changed the way of looking at the documentation and protection of CH. Moreover, Virtual Heritage, blends cultural data with virtual technologies and constitutes the future of Cultural/Religious Heritage documentation and enhancement in a modern and attractive direction.

*5.2 Digital Twins as a future trend for Cultural/Religious Heritage*

Someone may think various potentials that Mixed Reality technologies offer in the CH field, especially nowadays, that the platforms of Augmented Reality, the technological testbeds (e.g., MR headsets, smart glasses) and the networks (i.e. 5G/6G) are evolving with a tremendous rate. Applications cases include Virtual Museums, Virtual Tours and Augmented Reality expeditions in open cultural spaces (e.g., archaeological sites). In addition to the aforementioned considerations, the expanding concept of Digital Twins holds significant potential for efficient applications in cultural and religious heritage. This includes enhanced documentation, maintenance, reproduction, reconstruction, promotion, and project management. For instance, the Digital Twins framework allows for the simulation and virtual testing of material interactions with chemical products before undertaking actual invasive procedures, mitigating risks. It also enables real-time detection of vulnerabilities and risks associated with actual monuments. Therefore, the implementation of Digital Twins in cultural/religious heritage could: a) Establish intelligent systems that enable all involved parties to access heritage-related information anytime, anywhere. b) Integrate and connect various types of assets to form a digital heritage asset portfolio, enhancing access, management, sharing, and restoration efforts. c) Create a highly flexible, high-performance system capable of delivering easily understood information about monuments directly to stakeholders (e.g., Ephorate of Antiquities, Governmental organizations) and d) Proactively address the risks and dangers posed by climate change through informed decision-making and intervention strategies.

**Acknowledgments:** This research is partially supported by the project entitled: "Promotion and Dissemination of Cultural and Natural Heritage Through Development and Strengthening Religious Tourism in the Islands of Greece and Cyprus" referred as "RECULT" and is co-funded by the European Regional Development Fund (ERDF) and by national funds of Greece and Cyprus, under the Cooperation Programme "INTERREG V-A Greece-Cyprus 2014-2020 (MIS 5035557)". Access to the digital survey, products and deliverables is restricted only to the appropriate authorities and stakeholders.